\documentclass{article}

\usepackage{arxiv}

\usepackage[utf8]{inputenc} % allow utf-8 input
\usepackage[T1]{fontenc}    % use 8-bit T1 fonts
\usepackage{hyperref}       % hyperlinks
\usepackage{url}            % simple URL typesetting
\usepackage{booktabs}       % professional-quality tables
\usepackage{amsfonts}       % blackboard math symbols
\usepackage{nicefrac}       % compact symbols for 1/2, etc.
\usepackage{microtype}      % microtypography
\usepackage{graphicx}
\usepackage{natbib}
\usepackage{doi}
\usepackage[ruled]{algorithm2e}
\usepackage{makecell}
\usepackage{indentfirst}
\usepackage{graphicx}
\usepackage{txfonts}
\title{Deep-learning Assisted Extraction of Fluid Velocity from Scalar Signal Transport in a Shallow Microfluidic Channel}

\author{Xiao Zeng\\
	School of Optoelectronic Engineering \\and Instrumentation Science\\
	Dalian University of Technology\\
	Dalian 116024, Liaoning Province, P. R. China\\
	\texttt{zengxiao97@mail.dlut.edu.cn}\\
	\And
	Chundong Xue\thanks{Correspondence: Kairong Qin (krqin@dlut.edu.cn) and Chundong Xue (xuechundong@dlut.edu.cn)}\\
	School of Optoelectronic Engineering \\and Instrumentation Science\\
	Dalian University of Technology\\
	Dalian 116024, Liaoning Province, P. R. China\\
	\texttt{xuechundong@dlut.edu.cn}\\
	\And
	Kejie Chen\\
	School of Optoelectronic Engineering \\and Instrumentation Science\\
	Dalian University of Technology\\
	Dalian 116024, Liaoning Province, P. R. China\\
	\texttt{ckj@dlut.edu.cn}\\
	\And
	Yongjiang Li\\
	School of Optoelectronic Engineering \\and Instrumentation Science\\
	Dalian University of Technology\\
	Dalian 116024, Liaoning Province, P. R. China\\
	\texttt{yongjiangli@dlut.edu.cn}\\
	\And
	Kai-Rong Qin$^{*}$\\
	School of Optoelectronic Engineering \\and Instrumentation Science\\
	Dalian University of Technology\\
	Dalian 116024, Liaoning Province, P. R. China\\
	\texttt{krqin@dlut.edu.cn}
}

\date{}
\renewcommand{\shorttitle}{}
\hypersetup{
pdftitle={Deep-learning Assisted Extraction of Fluid Velocity from Scalar Signal Transport in a Shallow Microfluidic Channel},
pdfsubject={},
pdfauthor={Xiao Zeng},
pdfkeywords={Shallow Microfluidic Channel, Fluid Velocity Measurement, Scalar Image Velocimetry, Spatiotemporal Concentration Field, Deep Neural Networks},
}
\begin{document}
% ??? 修改title格式，作者和单位信息分开的形式。
\maketitle

\begin{abstract}
Precise measurement of flow velocity in microfluidic channels is of importance in microfluidic applications, such as quantitative chemical analysis, sample preparation and drug synthesis. However, simple approaches for quickly and precisely measuring the flow velocity in microchannels are still lacking. Herein, we propose a deep neural networks assisted scalar image velocimetry (DNN-SIV) for quick and precise extraction of fluid velocity in a shallow microfluidic channel with a high aspect ratio, which is a basic geometry for cell culture, from a dye concentration field with spatiotemporal gradients. DNN-SIV is built on physics-informed neural networks and residual neural networks that integrate data of scalar field and physics laws to determine the velocity in the height direction. The underlying enforcing physics laws are derived from the Navier–Stokes equation and the scalar transport equation. Apart from this, dynamic concentration boundary condition is adopted to improve the velocity measurement of laminar flow with small Reynolds Number in microchannels. The proposed DNN-SIV is validated and analyzed by numerical simulations. Compared to integral minimization algorithm used in conventional SIV, DNN-SIV is robust to noise in the measured scalar field and more efficiently allowing real-time flow visualization. Furthermore, the fundamental significance of rational construction of concentration field in microchannels is also underscored. The proposed DNN-SIV in this paper is agnostic to initial and boundary conditions that can be a promising velocity measurement approach for many potential applications in microfluidic chips.
\end{abstract}


\section{Introduction}
Microfluidic chip, a miniaturized device that integrates complex functions into a single chip with various microchannels, makes traditional serial processing and analysis in chemistry and biology be easily performed by manipulating the fluid at the microscale. It has several compelling advantages compared to traditional methods, such as little reagent consumption, faster analysis and response time, multi-functional integration and massive parallelization \citep{whitesides2006origins,stone2004engineering,nge2013advances}. With the widespread application of microfluidic technology in cell culture \citep{halldorsson2015advantages,sackmann2014present,shields2015microfluidic}, organs-on-chips \citep{esch2015organs,wu2020organ,wilmer2016kidney}, and many other exciting applications \citep{shang2017emerging,lam2005depthwise,doufene2019microfluidic}, the quantitative characterization of fluid in microchannels becomes more and more important. The shallow microchannel with a high aspect ratio (the height is much smaller than the other two dimensions) is one of the most common structure of microchannels especially for cell culture in microfluidic chips \citep{lane2012parallel,mehling2014microfluidic,van2009microfluidic}. In the shallow microchannel, the average flow velocity is legitimately used as the most important parameter to characterize the flow fields \citep{nauman1999quantitative,bacabac2005dynamic,wong2016parallel}. Thus precise measurement of the flow velocity in the microchannel is critical for the successful deployment of the microfluidic system \citep{yang2010manipulation,ong2008fundamental}.\par
A variety of velocimetry techniques have developed to quantitatively characterize the fluid velocity profile since the late 1900s \citep{sinton2004microscale,williams2010advances,wereley2010recent}. The most common velocimetry of the microfluidic flow are microscopic Particle Imaging Velocimetry (micro-PIV) and microscopic Particle Tracking Velocimetry (micro-PTV). The PIV method obtains the fluid velocity by measuring the displacement of the tracer particles, which are artificially introduced into the fluid and follow the fluid motion \citep{meinhart1999piv,lindken2009micro,westerweel2013particle}. In contrast, the PTV method directly measures and analyzes the movement of individual tracer particles, thus avoiding the Brownian motion of PIV and often achieves higher spatial accuracy \citep{adamczyk19882,kreizer2010real}. However, such particulate-based methods often bring about relatively large measurement errors when they are applied in shallow microchannels with a high aspect ratio \citep{wang2009measurement,williams2010advances}. The micron-sized tracers may interact with the upper and lower walls of the microchannels that leads to the poor flow following feature and a low measurement accuracy.\par
Scalar Imaging Velocimetry (SIV) is another kind of method that infers the fluid velocity from a conserved scalar field \citep{su1996scalar1,su1996scalar2}. It is developed based on the continuous mass transfer rather than the movement of individual particles. Thus, it avoids the interference of the flow field and following feature error caused by the size effect. Different from the optical flow velocimetry \citep{chen2015optical,kucukal2021blood}, SIV considers the diffusion effect during the mass transfer to make velocity measurement more accurate. The prerequisite is that the scalar field must contain more information than the velocity field. The time scalar and the dissipation length scalar of dynamic scalar field must be smaller than the corresponding scalar of dynamic velocity field in the interested flow field \citep{gillissen2018space}. Hence, the SIV method can provide an accurate estimation of the large-scale turbulent flow with high Reynolds Number where the convection is dominating \citep{gillissen2018space,heitz2010variational}. Meanwhile, SIV has also been applied frequently for experimentally visualizing the turbulence of geophysical flow in meteorology, climatology, and oceanography \citep{kalnay2003atmospheric,papadakis2008variational}.\par
However, it is usually hard to accurately estimate the fluid flow in microfluidic channels using the original SIV method because the microfluidic flow at low Reynolds Numbers $(Re<1)$ is not convection-dominated but rather convection-diffusion-cooperated. The spatiotemporal gradient of the scalar field is modulated by the coupling effect of the convection and diffusion. This coupling effect becomes more complex when diffusion is more dominated. Moreover, it has been reported \citep{li2013transport,chen2016shaped} that the transport of dynamic scalar signals in a shallow microfluidic channel can be modulated by different flow behaviors, such as the pulsatile flow. The amplitude and frequency of concentration signal are nonlinearly modulated by the amplitude and the frequency of the pulsatile flow \citep{li2018transmission}. How the interplay between the concentration signal and flow velocity affects the velocity reconstruction from scalar signal transport in the shallow straight microchannels remains elusive. In fact, to the best of our knowledge, the SIV method has rarely been successfully used for the characterization of fluid in microfluidic chips.\par
The common implementation of SIV is to apply an integral minimization algorithm of a cost function that minimizes weighted residuals of conserved scalar transport equation combining continuity condition and conserved fluid momentum \citep{heitz2010variational,papadakis2008variational}. However, quite a few studies have reported \citep{burman2020stability} that the accuracy of the velocity reconstruction by minimizing the cost function is sensitive to noisy data of scalar fields and has poor robustness even in flows at a high Reynolds Number. It is likely that the effects of noise on SIV would be sharply augmented for microscale flows at a small Reynolds Number. New algorithms need to be proposed to address the issues of noise sensitivity and robustness. While data-driven model has being gradually emerged in the fluid mechanics \citep{kou2021data,bezgin2021data}, physics-informed deep learning framework (PINN) has been widely used in solving the forward and inverse problems in fluid mechanics \citep{chen2021learning,raissi2020hidden,raissi2019physics}. The deep neural networks (DNN) have been reported to be able to break the limits of accuracy and extensibility in modeling of fluid mechanics, showing better performance in accuracy, stability and prediction power \citep{barwey2019using,cai2021artificial}.\par
To address the aforementioned issues, by integrating the SIV concept and deep learning, we propose a deep neural networks assisted scalar image velocimetry (DNN-SIV) for quick and precise extraction of the average fluid velocity from the spatiotemporal concentration gradients in a shallow microfluidic channel. In order to improve velocimetry in microchannel, multiple dynamic concentration boundary conditions are applied at the inlet of the microchannel and the spatiotemporal profiles are generated by regulating the transport of the dynamic concentration signals. Based on the principles of fluid mechanics and the scalar transport, the proposed approach is validated via direct numerical simulation studies. In addition, the effects of influential factors including the scalar signal transport and fluid flow characteristics are analyzed.

\section{Method}
\subsection{Mathematical model}
\begin{figure}
	\centering
	\includegraphics[width=0.8\textwidth]{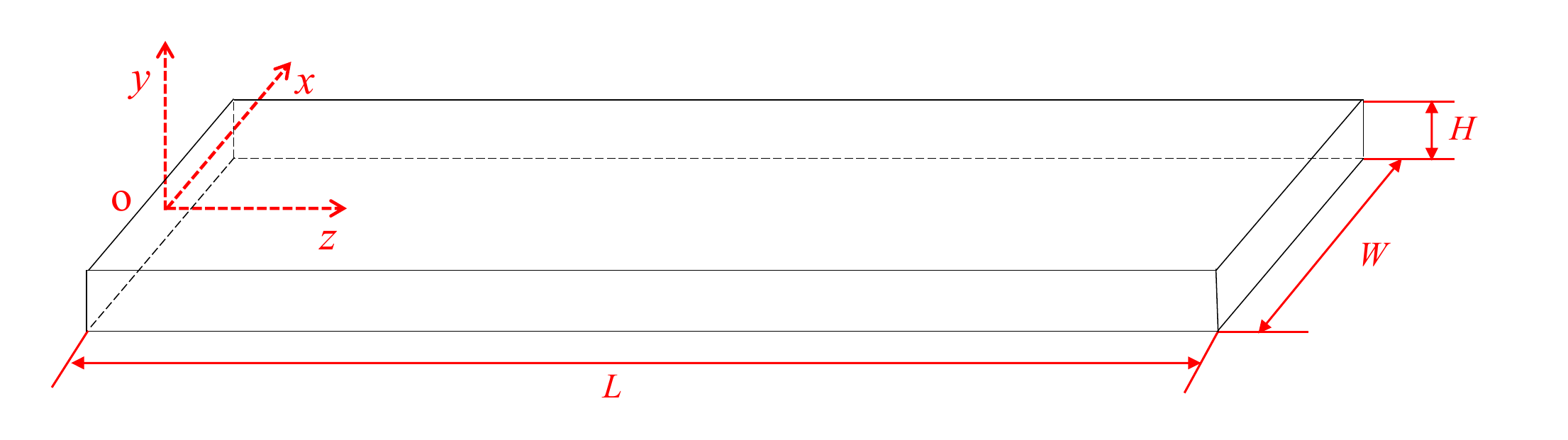}
	\caption{Schematic and geometry of a straight shallow microchannel with a high aspect ratio in a microfluidic chip.}
	\label{fig:fig1}
\end{figure}

The typical shape of the shallow microchannel considered in the present study is a straight rectangle with a high aspect ratio as shown in Fig \ref{fig:fig1}., which is one of the most common structures for cell culture in microfluidic chips \citep{lane2012parallel,mehling2014microfluidic,van2009microfluidic}. The channel height $H$ is much smaller than the channel width $W$, i.e., $H/W\ll1$. The fluid in the microchannel is assumed to be incompressible, viscous and Newtonian. In most microfluidic applications, the fluid flow in the microchannels has a small Reynolds Number and a small Womersley Number. Thus, the flow field can be described by the simplified Navier-Stokes equation as follows \citep{li2013transport},
\begin{equation}
	\frac{\partial u}{\partial t}=-\frac{1}{\partial p}\frac{\partial p}{\partial z}+\frac{\mu}{\rho}\frac{\partial^2u}{\partial y^2},
\end{equation}	
where $u$ is the velocity in $z$-direction, $p$ is the pressure, $\rho$ and $\mu$ are the density and the dynamic viscosity of the fluid, respectively. Under the assumption of quasi-steady flow, the velocity profile $u\left(y,t\right)$ can be derived as follows \citep{li2018transmission},
\begin{equation}
	u\left(y,t\right)=\frac{3}{2}\left[1-\left(\frac{2y}{H}\right)^2\right]\bar{u}\left(t\right),
\end{equation}
\begin{equation}
	\bar{u}\left(t\right)=\frac{Q\left(t\right)}{WH},
	\label{Eq:3}
\end{equation}
where $\bar{u}\left(t\right)$ and $Q\left(t\right)$ denote the average velocity along the channel height and the flow rate, respectively.\par

If there is some substance, such as fluorescent dyes or fluorescent protein molecules, inside the fluid, the transport of substance is governed by the convection-diffusion equation,
\begin{equation}
	\frac{\partial\mathbf{c}}{\partial t}+u\frac{\partial\mathbf{c}}{\partial z}=D\nabla^2\mathbf{c},
	\label{Eq:4}
\end{equation}
where $\mathbf{c}$ is the substance concentration and $D$ is the substance diffusivity. Since $H/W\ll1$, the substance can be quickly and uniformly distributed along the channel height direction (\emph{y}-direction). The concentration in \emph{y}-direction is thus assumed to be homogeneous. The average concentration along the channel height, $\bar{\mathbf{c}}\left(x,z,t\right)$, can be defined as,
\begin{equation}
	\bar{\mathbf{c}}\left(x,z,t\right)=\frac{1}{H}\int_{-H/2}^{H/2}\mathbf{c}\left(x,y,z,t\right)dy,
\end{equation}
which satisfies Taylor-Aris dispersion equation as follows \citep{li2018transmission},
\begin{equation}
	\frac{\partial\bar{\mathbf{c}}}{\partial t}+\bar{u}	\frac{\partial\bar{\mathbf{c}}}{\partial z}=D_{eff}\frac{\partial^2\bar{\mathbf{c}}}{\partial z^2},
	\label{Eq:6}
\end{equation}
\begin{equation}
	\bar{\mathbf{c}}\vert_{z=0}=\bar{\mathbf{c}}_0\left(t\right),
	\label{Eq:7}
\end{equation}
\begin{equation}
	\frac{\partial\bar{\mathbf{c}}\left(z,t\right)}{\partial t}\vert_{z=0}=0,
	\label{Eq:8}
\end{equation}
where $D_{eff}$ is the effective diffusivity coefficient, that is usually expressed as,
\begin{equation}
	D_{eff}=D\left[1+\frac{1}{210}\left(\frac{\bar{u}H}{D}\right)^2\right],
	\label{Eq:9}
\end{equation}
where $\bar{\mathbf{c}}_0\left(t\right)$ is the concentration boundary condition at the inlet. And the periodic concentration signal $\bar{\mathbf{c}}_0\left(t\right)$ is considered in this study expressed as,
\begin{equation}
	\label{Eq:10}
	\bar{\mathbf{c}}_0\left(t\right)=C_0\left(1+\epsilon_c\sin\left(\omega_ct\right)\right),
\end{equation}
where $C_0$ is the time-averaging concentration which is a constant, $\epsilon_c$ is the amplitude of the fluctuating concentration signal, and $\omega_c=2\pi f_c$ is the angular frequency corresponding to the frequency $f_c$. If the temporal and spatial gradients of the concentration in the microchannel, i.e., $\frac{\partial \bar{\mathbf{c}}}{\partial t}$, $\frac{\partial\bar{\mathbf{c}}}{\partial z}$, and $\frac{\partial^2\bar{\mathbf{c}}}{\partial z^2}$ are given, the average velocity $\bar{u}$ can be solved based on Eq.(\ref{Eq:6}).

\subsection{DNN-assisted SIV method}
Conventional SIV method is mainly applied in dense velocity field and turbulent flow measurement in experimental fluid mechanics, where the flow is convection-dominated and the Reynolds Number is high \citep{feng2007simultaneous,corpetti2009pressure}. Conventional SIV extracts velocity from a scalar field by applying an integral minimization algorithm of the cost function that penalizes the deviation from the scalar transport equation under constraint conditions. In a straight shallow microchannel, the Reynolds Number is small and both the convection and diffusion play significant roles \citep{li2018transmission}. Inspired by the minimization algorithm of conventional SIV, the average velocity $\bar{u}$ can be computed by minimizing the cost function, $\mathcal{J}\left(\hat{u}\right)$, which is constructed by an integral of the one-dimensional Taylor-Aris dispersion equation (Eq.(\ref{Eq:6})) as,
\begin{equation}
	\mathcal{J}\left(\hat{u}\right)=\int_\Omega\left(\frac{\partial\bar{\mathbf{c}} }{\partial t}+\hat{u}\frac{\partial\bar{\mathbf{c}} }{\partial z}-D_{eff}\frac{\partial^2\bar{\mathbf{c}} }{\partial z^2}\right)dz,
	\label{Eq:11}
\end{equation}
where $\Omega$ denotes the region of interest and $\hat{u}$ denotes the height-averaging velocity extracted from the scalar concentration field. It is worth noted that the cost function (Eq. (\ref{Eq:11})) needs no additional constraint of continuity equation since the velocity in a straight shallow microchannel is only time dependent.\par
\begin{figure}
	\centering
	\includegraphics[width=0.9\textwidth]{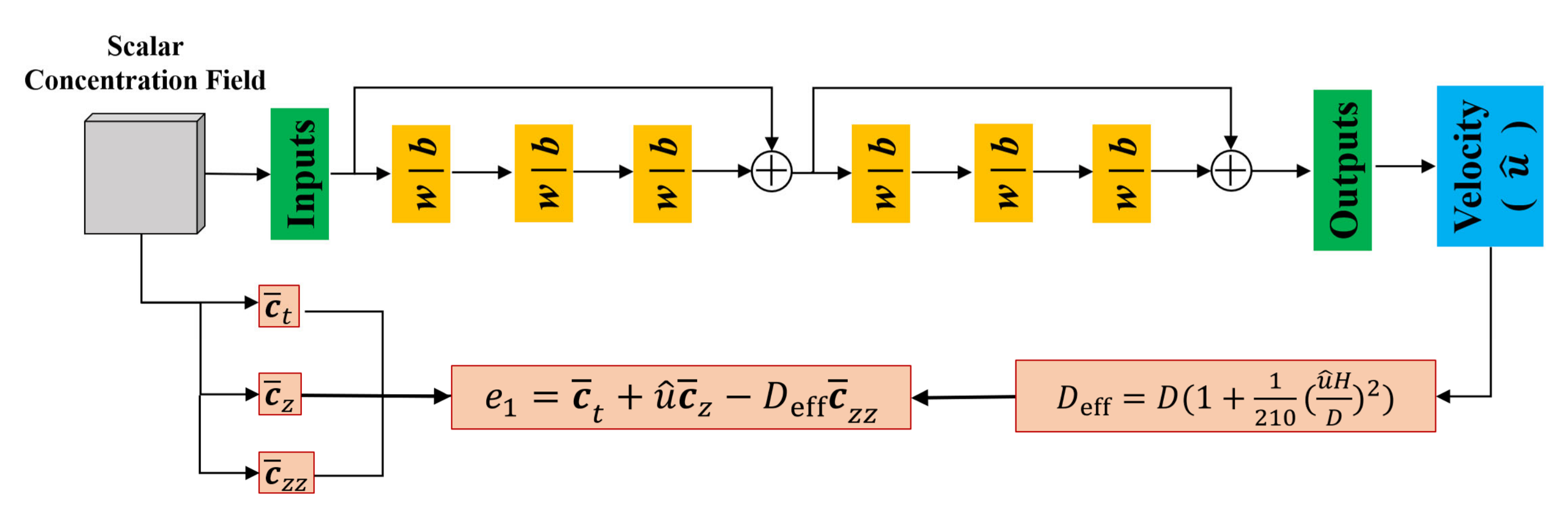}
	\caption{Schematic of physics-informed residual networks with fully connected layers for extraction of height-averaging velocity from the scalar concentration field. The input is the scalar concentration field at successive moments and output is the height-averaging velocity. The residual of physical constraint attached in neural networks is derived from Taylor-Aris dispersion equation (Eq. \ref{Eq:6}).}
	\label{fig:fig2}
\end{figure}

As stated in the introduction, the accuracy of the velocity extraction by the conventional SIV approach is sensitive to the noisy data of the scalar fields and has poor robustness \citep{wallace2010measurement,burman2020stability}. To address this issue, a DNN framework integrating scalar transport equation is designed to map the scalar concentration field  to the height-averaging velocity. As depicted in Fig.\ref{fig:fig2}, a physics-informed residual networks framework integrating SIV is an acyclic cascade consisting of the physics-informed neural networks \citep{raissi2019physics} and the residual neural networks \citep{he2016deep}, of which the physics-informed neural networks deal with the physical constraints and the residual neural networks deal with the vanishing gradients and mitigates the degradation problem. We referred to this method as the DNN-assisted SIV method (DNN-SIV). The main idea of the proposed DNN model is to approximate the velocity extracted from the concentration field as,
\begin{equation}
	\label{Eq:12}
	\hat{u}\left(t_k\right)=f\left[\bar{\mathbf{c}}\left(t_k,t_{k+1}\right),D,H;\mathbf{\Theta}\right],
\end{equation}
where $t_k$ and $t_{k+1}$ are two successive moments,  $\hat{u}\left(t_k\right)$ is the averaging velocity predicted by the DNN at moment $t_k$, $\bar{\mathbf{c}}\left(t_k,t_{k+1}\right)$ represents the normalized concentration in \emph{z}-direction at moment $t_k$ and $t_{k+1}$, $D$ and $H$ are the diffusivity and height of the microchannel, respectively. $\mathbf{\Theta}\left(\mathbf{W},\mathbf{b}\right)$ is weights and biases of the DNN model. The DNN models is composed of a fully-connected neural network with six hidden layers. For the fully-connected networks, the relationship between \emph{i}-\emph{th} layer and the (\emph{i}+1)-\emph{th} layer can be expressed as,
\begin{equation}
	\sigma\left(\mathbf{W_ix_i+b_i}\right)\mapsto \mathbf{x_{i+1}},
\end{equation}
where $\mathbf{W_i}$ and $\mathbf{b_i}$ are the weights and biases, $\mathbf{x}_i$ is input vector of \emph{i}-\emph{th} layer and $\sigma\left(\cdot\right)$ denotes the Rectified Linear Unit (ReLU) activation function.\par
In order to infer the averaging velocity, an additional constraint with the underlying physics is encoded into the DNN. Specifically, we aim to minimize the residual of the governing equation in the networks, which can be thought of as enforcing for the feed-forward networks output. The residual of the scalar transport equation (Eq. \ref{Eq:6}) is defined as,
\begin{equation}
	\label{Eq:14}
	e_1=\bar{\mathbf{c}}_t+\hat{u}\bar{\mathbf{c}}_z-D\left(1+\frac{1}{210}\left(\frac{\hat{u}H}{D}^2\right)\right)\bar{\mathbf{c}}_{zz},
\end{equation}
where the subscripts represent the derivatives of the corresponding quantities, $\bar{\mathbf{c}}_t$, $\bar{\mathbf{c}}_z$, and $\bar{\mathbf{c}}_{zz}$. Specially, $\bar{\mathbf{c}}_t=\frac{d\bar{\mathbf{c}}}{dt}$, $\bar{\mathbf{c}}_z=\frac{\partial \bar{\mathbf{c}}}{\partial z}$, and $\bar{\mathbf{c}}_{zz}=\frac{\partial^2 \bar{\mathbf{c}}}{\partial z^2}$. By integrating the mean square error and residual of the scalar transport equation, the loss function, $\mathcal{L}$, for training the DNN is then defined as:
\begin{equation}
	\mathcal{L}=\lambda_f\mathcal{L}_f+\lambda_{SIV}\mathcal{L}_{SIV},
	\label{Eq:15}
\end{equation}
\begin{equation}
	\mathcal{L}_f=\sum\Vert\hat{u}\left(t_k\right)-\bar{u}\left(t_k\right)\Vert_2,
	\label{Eq:16}
\end{equation}
\begin{equation}
	\mathcal{L}_{SIV}=\sum\Vert\ e_1\left(\hat{u},\bar{\mathbf{c}},D,H,\right)\Vert_2,
	\label{Eq:17}
\end{equation}
where $\lambda_f$ and $\lambda_{SIV}$ are penalty weighting factors, $\Vert\quad\Vert_2$ is the Euclidean norm.
\par
We train the DNN model by minimizing the loss function (Eq. \ref{Eq:15}) composed of a data mismatch term (Eq. \ref{Eq:16}) and residual term (Eq. \ref{Eq:17}) associated with the scalar transport equation. Instead of directly computing the averageing velocity by minimizing the cost function (Eq. \ref{Eq:11}) during each experiment, the DNN-SIV method firstly trains a surrogate model (Eq. \ref{Eq:12}) by minimizing the loss functions (Eq. \ref{Eq:15}), and then use the measured the spatiotemporal concentration to quickly predict the height-averaging velocity. \par
The parameters of the neural networks are initialized using the Xavier scheme \citep{glorot2010understanding}, and then optimized until convergence. Firstly, forward propagation is used to obtain the prediction of the velocity $\hat{u}$. Then the cost function is computed to acquire the Euclidean distance between the vectors and the residual of the scalar transport equation. Finally, the Adam algorithm \citep{kingma2014adam} is adopted to compute the cost function gradients and the weights and biases of networks are updated by the backpropagation algorithm. The total training procedure is summarized in Algorithm \ref{alg:algorithm1}. When the optimal parameters $\mathbf{\Theta}$ are obtained after training, the velocity can be inferred very fast by feeding the scalar concentration field data to the DNN model.
% ？？？ 修改算法框宽度并居中
\begin{algorithm}
	\caption{DNN Training Scheme.}
	\label{alg:algorithm1}
	\KwIn{Training dataset $\left(\bar{\mathbf{c}}, \bar{u}\right)$, Shape parameters $\left(D,H\right)$, \\Learning rate $\eta$, and Penalty weighting factors $\left(\lambda_{f},\lambda_{SIV}\right)$.}	\KwSty{Initialization:} Randomly initialize $\mathbf{\Theta}$\;
	\While{not convergent}{
		$\mathcal{L}_f\gets\mathbf{FORWARD}\left(\bar{\mathbf{c}},\bar{u};\mathbf{\Theta}\right)$\;
		$\mathcal{L}_{SIV}\gets\mathbf{FORWARD}\left(\bar{\mathbf{c}},\hat{u},D,H;\mathbf{\Theta}\right)$\;
		$\mathcal{L}\left(\lambda_{f},\lambda_{SIV}\right)=\lambda_{f}\mathcal{L}_f+\lambda_{SIV}\mathcal{L}_{SIV}$\;
		$\mathbf{\Theta}\gets\mathbf{BACKWARD}\left(\mathcal{L};\mathbf{\Theta}\right)$\;
	}
	\KwOut{Return the optimal $\mathbf{\Theta}$ }
\end{algorithm}

\subsection{Numerical simulation method}
\label{Simulation}
The dataset of the concentration field and the velocity profile for the validation of the above method is obtained from forward numerical simulation. Briefly, given the height-averaging velocity $\bar{u}$ (Eq. \ref{Eq:3}) as well as the initial conditions and boundary conditions (Eqs. \ref{Eq:7} -- \ref{Eq:10}), the spatial and temporal evolution of the concentration in a straight shallow channel is solved numerically. The computer code based on the finite difference methods is developed to acquire the numerical solution of spatial and temporal gradients of concentration, including a central difference approximation in spatial gradient and a forward difference approximation in time gradient. \par
In the forward numerical simulations, we discretize the channel length in spatial step $\Delta z$ along the \emph{z}-direction. The spatial grid points are denoted as $z_m(m = 1\cdots M)$. Time is discretized in time step $\Delta t$ and the discretized moment is denoted as $t_k (k = 1 \cdots N)$. Dataset is constructed through normalizing the simulation data of concentration field and recording the corresponding velocities.\par
To validate the stability and robustness of the proposed method, white noise is added in the dataset of concentration field,
\begin{equation}
	\bar{\mathbf{c}}_{noise}\left(t_k\right)=\bar{\mathbf{c}}\left(t_k\right)\left(1+\alpha W\left(M\right)\right),
\end{equation}
where $\bar{\mathbf{c}}_{noise}$ denotes the concentration field added with noise, $\bar{\mathbf{c}}$ denotes the original concentration field without noise, $\alpha$ is a coefficient between 0 and 1 representing the intensity of white noise and $W(M)$ is a $M$ uncorrelated random variables with zero mean and finite variance approximating discrete-time white noise.\par
In order to quantitatively evaluate the accuracy of the velocity prediction by DNN-SIV, the absolute percentage error (APE) and mean absolute percentage error (MAPE) are used, which are defined as,
\begin{equation}
	APE=\left|\frac{T_i-P_i}{T_i}\right|,
\end{equation}
\begin{equation}
	MAPE=\frac{1}{N}\sum_{i=1}^{N}\left|\frac{T_i-P_i}{T_i}\right|,
\end{equation}
where $N$ is the number of data points, $T_i$ and $P_i$ are the true and prediction values of the \emph{i}-th data point, respectively.\par
All aforementioned algorithms and calculations were built in MATLAB (The Math Works R2020b, Inc). The default values of the parameters used in baseline simulations are listed in Table \ref{tab:table1}. Unless otherwise specified, these values are used for all the simulations.

\begin{table}
	\caption{Default values of the straight shallow microchannel parameters used in numerical simulations.}
	\centering
	\begin{tabular}{ll}
		\toprule
		Parameters & Values\\
		\midrule
		$L\left(x-direction\right)$ & $30~mm$   \\
		$W\left(y-direction\right)$ & $3~mm$    \\
		$H\left(z-direction\right)$ & $0.15~mm$ \\
		$\Delta x$                  & $0.5~mm$  \\
		$\Delta y$                  & $0.5~mm$  \\
		$\Delta t$                  & $0.001~s$  \\
		$D$                         & $8.3\times10^{-11}m^2/s$  \\
		$C_0$                  		& $1~mmol/m^3$  \\
		$\epsilon_c$                & $1$  \\
		$f_c$		                & $2$ Hz  \\
		\bottomrule
	\end{tabular}
	\label{tab:table1}
\end{table}

\section{Results}
\subsection{Velocity Extraction}
\paragraph{A. Noise-free data of concentration field}
\begin{figure}
	\centering
	\includegraphics[width=0.6\textwidth]{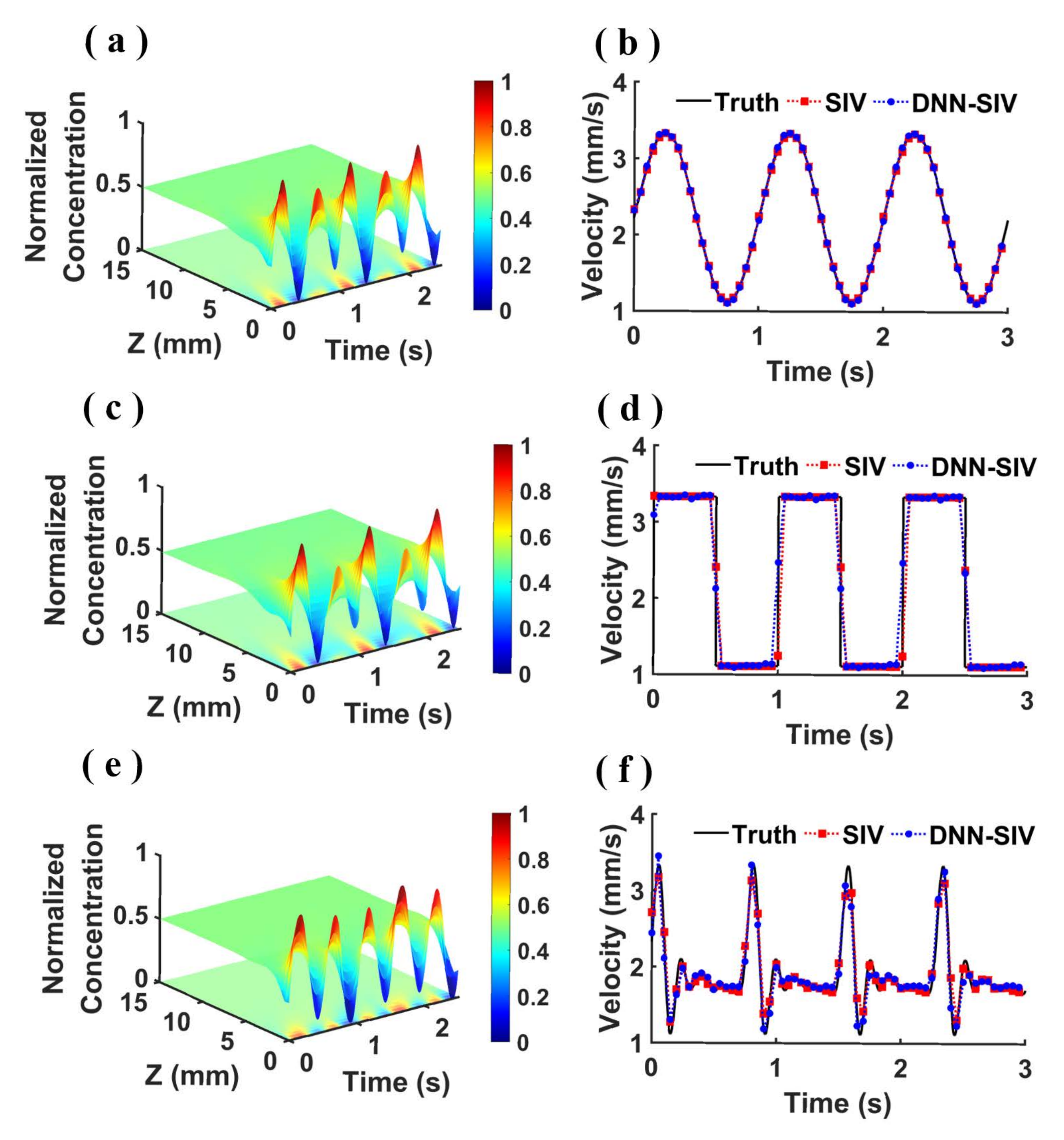}
	\caption{ Comparison of different flow waveforms extracted from the noise-free data of scalar concentration field based on conventional SIV method and DNN-SIV. (a), (c), and (e) are distributions of normalized spatiotemporal concentration generated by different flow waveforms and same dynamic concentration boundary conditions as presented in Table \ref{tab:table1}. (b), (d), and (f) are the extracted hegiht-averaging velocity profiles of sinusoidal, square, and physiological blood-flow-like waveform, respectively. Black solid line denotes true velocity. Red square dashed line and blue point dashed line are the extracted velocities using conventional SIV method and DNN-SIV method respectively.}
	\label{fig:fig3}
\end{figure}

To validate the DNN-SIV proposed in the present study, we first extract the velocity from a noise-free concentration field in Figure \ref{fig:fig3}. Three spatiotemporal profiles of the normalized noise-free concentration field are simulated in the microchannel and used as the datasets, as shown in Figure \ref{fig:fig3}(a), (c) and (e), respectively. The corresponding pulsatile flow with three different waveforms, i.e., sinusoidal, square and physiological blood-flow-like waveform are referred as ``Truth'' in Figure \ref{fig:fig3}(b), (d), and (f), respectively. Both the integral minimization algorithm of the conventional SIV and the DNN-SIV are used and the corresponding results are denoted as ``SIV'' and ``DNN-SIV'', respectively. It is clear that the extracted velocity waveforms are very close to the true values, even for the case of physiological blood-flow-like waveform. This demonstrates the ability of our proposed DNN-SIV in extraction of flow velocity in the shallow microchannel.\par
In addition, for all the cases, no significant difference is observed between the velocity waveforms extracted using SIV and DNN-SIV methods. It is conceivable because the data of concentration field used here are noise-free. However, the measurement error of concentration field is inevitable in experiments. The performance of the two methods needs to be further evaluated and compared for the case of noisy concentration field.

\paragraph{B. Noisy data of concentration field}\par
\begin{figure}
	\centering
	\includegraphics[width=0.8\textwidth]{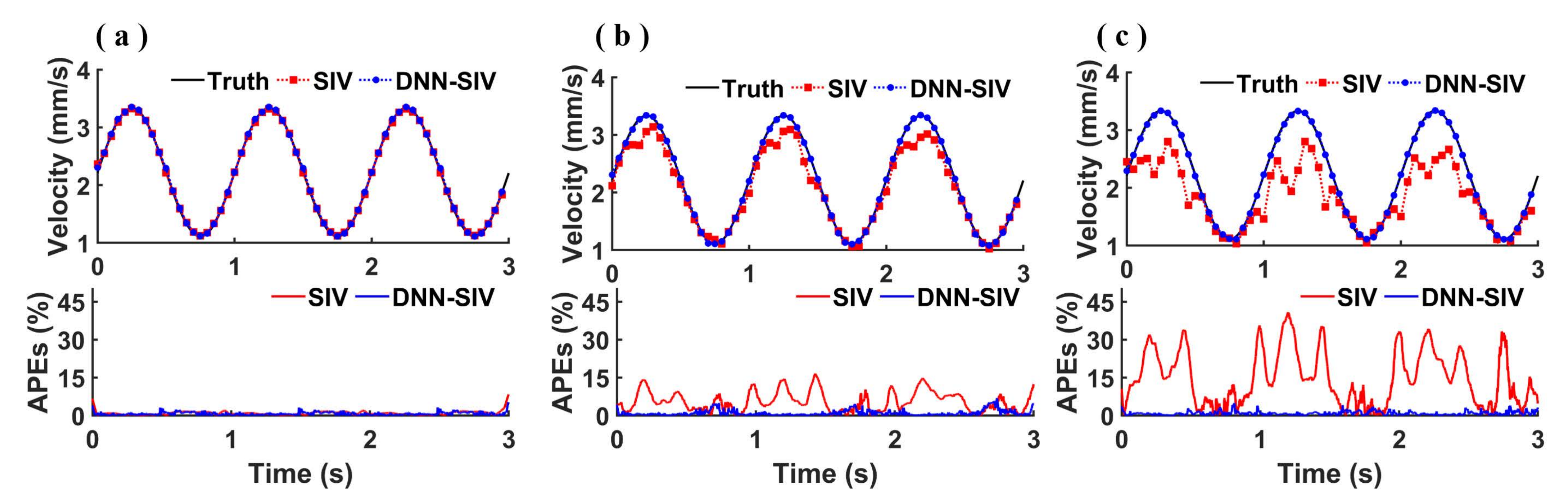}
	\caption{Effect of noise on the conventional SIV method and DNN-SIV method. (a), (b), and (c) are the velocity profiles and corresponding APEs. The velocity profiles in (a), (b), and (c) are extracted from scalar field data with different levels of white noise (0.1\%, 2\%, and 4\%) as stated in Sect. \ref{Simulation} respectively. Black solid line denotes true velocity. Red square dashed line and blue point dashed line denote conventional SIV method and DNN-SIV method respectively. The APEs of extracted velocity using convention SIV method (red solid line) and DNN-NN method (blue solid line) are presented below.}
	\label{fig:fig4}
\end{figure}

\begin{table}
	\caption{Time consumption comparison of conventional SIV method and DNN-SIV method.}
	\centering
	\begin{tabular}{ccc}
		\toprule
		Data Points & \makecell{Consumption time of \\conventional SIV method} &
		\makecell{Consumption time of \\DNN-SIV method} \\
		\midrule
		$1\times10^3$   & $1.23~s$  & $0.14~s$ \\
		$5\times10^3$   & $6.19~s$  & $0.31~s$ \\
		$1\times10^4$   & $12.25~s$ & $0.51~s$ \\
		\bottomrule
	\end{tabular}
	\label{tab:table2}
\end{table}

Figure \ref{fig:fig4} demonstrates the velocity extracted from the concentration field with different noisy levels. The noisy concentration filed are constructed by adding white noise to the idealized concentration filed as stated in Sect. \ref{Simulation}. The results for the case of three levels of white noise, i.e., 0.1\%, 1\%, and 4\%, are given in Figure \ref{fig:fig4}(a), (b) and (c), respectively. The corresponding APEs for two methods is then calculated and compared in bottom of Figure \ref{fig:fig4}. For simplicity, only the results for the sinusoidal pulsating flow are illustrated here since the sinusoidal waveform is the most basic waveform. The other periodic waveforms can be obtained by superimposing the sinusoidal waveform based on the Fourier transform. All the following results are illustrated using the sinusoidal pulsating flow. \par
It is evident from the Fig. \ref{fig:fig4} that the APE of conventional SIV method increases significantly as the noise grows (Fig. \ref{fig:fig4}(b) and (c)). The APE of the DNN-SIV method, in contrast, remains low at different levels of noise, showing the robustness of the DNN-SIV. Another advantage of DNN-SIV compared to conventional SIV method is time consumption. Although the training process of neural networks is fairly time-consuming, once the training is completed the efficiency of DNN-SIV method in velocity extraction has a huge improvement over conventional SIV method (Table. \ref{tab:table2}). The fast processing speed makes it possible to visualize the real-time flow velocity in the microchannel and extend its application domain to those with high response time demand.

\subsection{Effects of scalar signal transport and fluid flow characteristics}
\label{Sec:ScalarFLuid}
\paragraph{A. Scalar signal transport}
\begin{figure}
	\centering
	\includegraphics[width=0.8\textwidth]{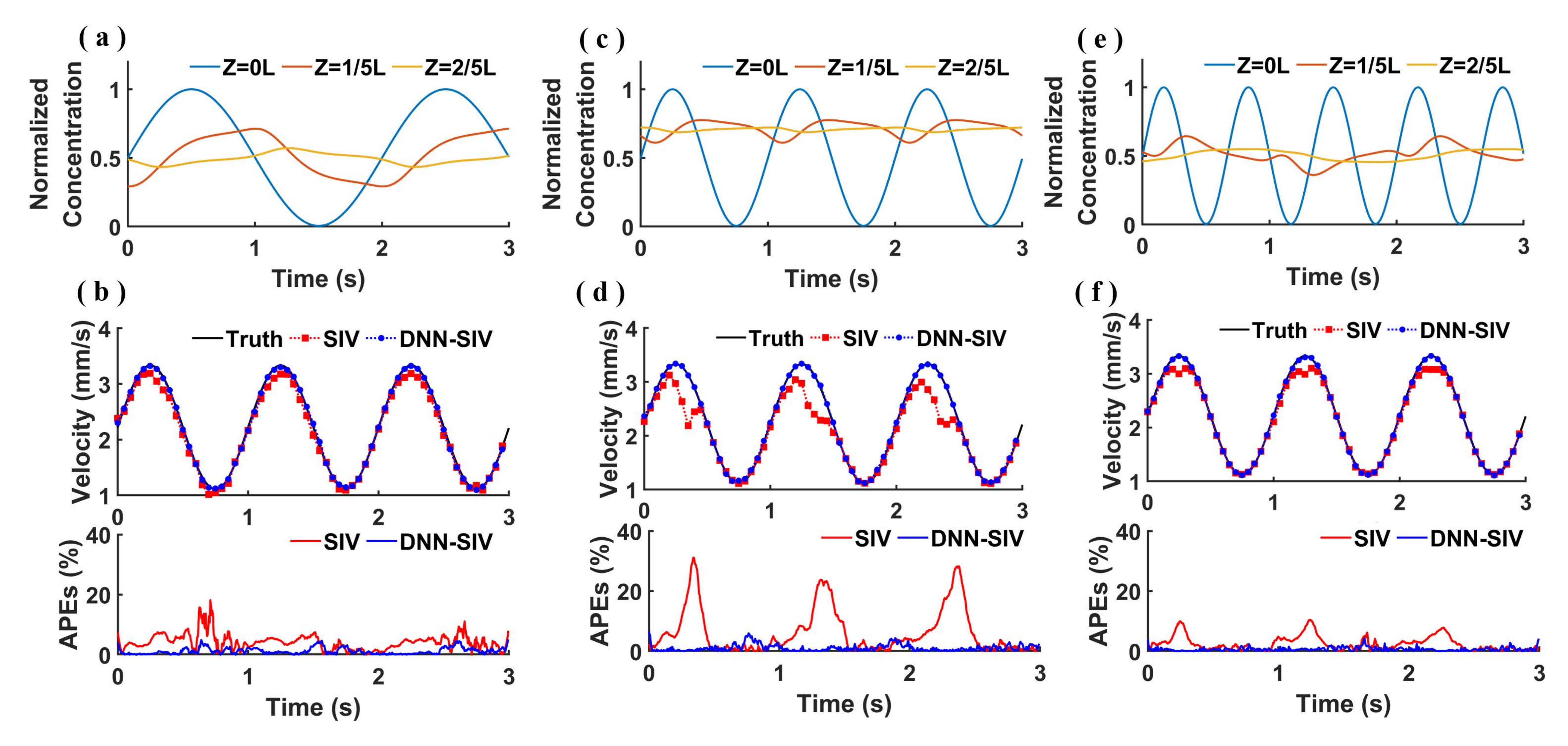}
	\caption{Effect of concentration signal transport under different boundary conditions of flow frequency and concentration frequency. (a), (b), and (c) show the concentration signal at three different locations of the straight shallow microchannel (Inlet: blue solid line; One-fifth from inlet: orange solid line; Two-fifth from inlet: yellow solid line). The corresponding flow frequency and concentration frequency in (a), (c), and (e) are 1Hz and 0.5Hz, 1Hz and 1Hz, and 1Hz and 1.5Hz, respectively. Those are for the case where flow frequency is less than or equal or bigger than concentration frequency at inlet. The corresponding true velocity profile (black solid line) and extracted velocity profile using conventional SIV method (red square dashed line) and DNN-SIV method (blue point dashed line) is shown in (b), (d), and (f). The APEs of extracted velocity using convention SIV method (red solid line) and DNN-NN method (blue solid line) are presented below.}
	\label{fig:fig5}
\end{figure}

\begin{figure}
	\centering
	\includegraphics[width=0.8\textwidth]{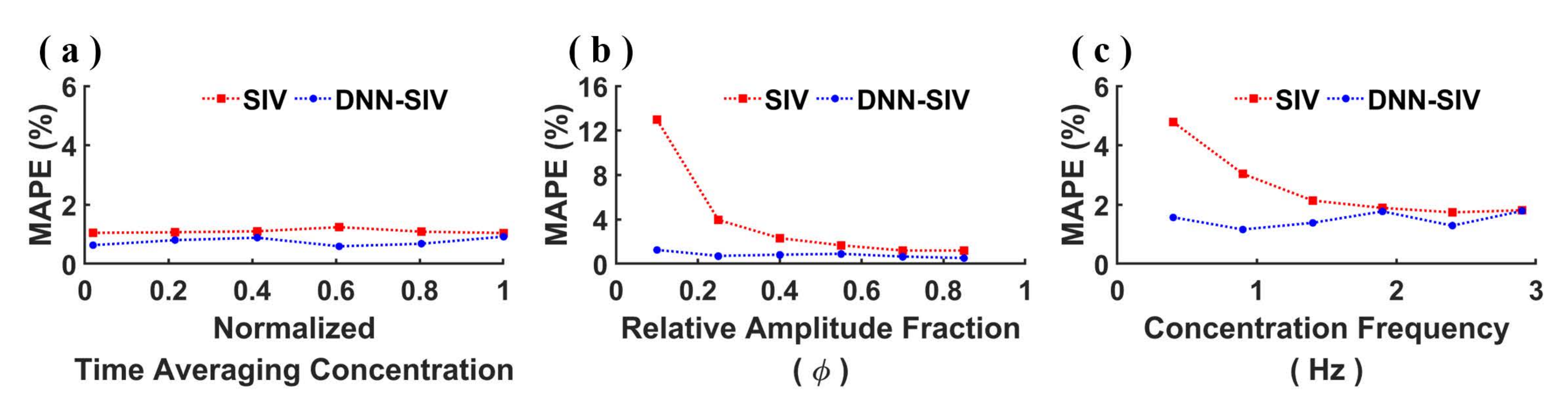}
	\caption{The effect of scalar concentration signal transport. The MAPE of three parameters are presented (N=3000), including normalized time averaging concentration (a), relative amplitude fraction (b), and concentration frequency (c). The red square dashed line denotes MAPE of conventional SIV method and the blue point dashed line denotes the MAPE of DNN-SIV method.}
	\label{fig:fig6}
\end{figure}

It should be underscored that the construction of concentration filed is of fundamental significance to our DNN-SIV approach. It is clearly shown in Fig \ref{fig:fig5} that different combination of flow and concentration frequency could generate quite different scalar signals in the microchannel, as shown in Fig \ref{fig:fig5}(a), (c), and (e). The APEs of velocity extracted from different scalar signals are significantly distinct, as shown in Fig \ref{fig:fig5}(b), (d), and (f). When the frequencies of the velocity and the concentration are equal, a flatter scalar signal is generated in the microchannel due to the low-pass filtering effect and nonlinear modulation \citep{li2018transmission}, resulting in a relatively larger APE compared to the other two cases. This reveals that, in addition to improving velocity extraction algorithms, the APEs of extracted velocity can be reduced by adjusting the velocimetry parameters of SIV. In the following, we analyze other factors including normalized time averaging concentration, relative amplitude fraction and concentration frequency that may affect velocity extraction.\par
The effect of three parameters with respect to scalar concentration signal transport has been shown in Fig \ref{fig:fig6}. The magnitude of normalized time averaging concentration seems to have no notable effect on the MAPE of extracted velocity as shown in Fig. \ref{fig:fig6}(a). It is reasonable that the velocity extraction using SIV method is related to the gradient of scalar field and independent of the absolute value. It is the same reason for the influence of the relative amplitude fraction on the MAPE as displayed in Fig. \ref{fig:fig6}(b). The larger relative amplitude fraction, the larger the gradient of the scalar field generated by concentration signal. Thus, the MAPE decreases with increasing relative amplitude fraction. As presented in Fig. \ref{fig:fig6}(c), the MAPE first decreases and then remains stable as the concentration frequency increases for a given flow frequency.

\paragraph{B. Fluid flow characteristics}
\begin{figure}
	\centering
	\includegraphics[width=0.8\textwidth]{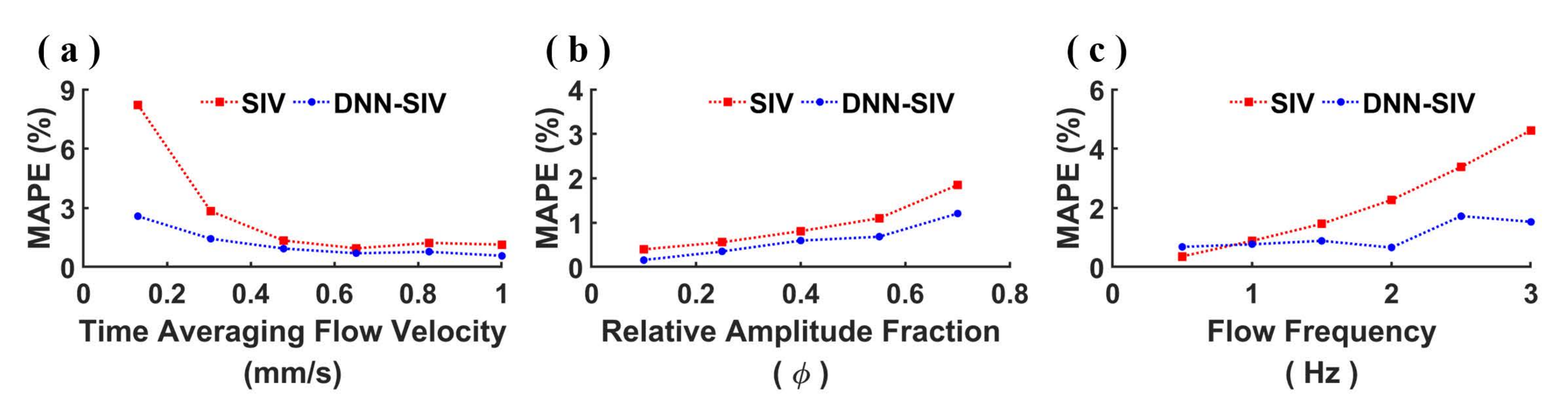}
	\caption{The effect of fluid flow characteristics. The MAPE of three parameters are presented (N=3000), including time averaging flow velocity (a), relative amplitude fraction (b), and flow frequency(c). The red square dashed line denotes MAPE of conventional SIV method and the blue point dashed line denotes the MAPE of DNN-SIV method.}
	\label{fig:fig7}
\end{figure}

The effects of three parameters with respect to fluid flow characteristics have been exhibited in Fig. \ref{fig:fig7}. As shown in Fig. \ref{fig:fig7}(a), MAPE first decreases rapidly and then gradually flatters as the flow rate increases. The MAPE of the extracted velocity is bigger at low flow velocity due to the reduction of the convective effect and the growing of diffusion effect. Besides, the relative amplitude fraction and the MAPE of extracted velocity are approximately linearly and positively correlated, as shown in Fig. \ref{fig:fig7}(b). The magnitude of the relative amplitude fraction represents the fluctuation range of the flow rate, and it is obvious that the larger the fluctuation ranges, the larger the extracted velocity error is. As for the influence of the flow frequency, it is shown in Fig \ref{fig:fig7}(c) that the MAPE rises quite fast as the flow frequency increases. Based on Fig. \ref{fig:fig6}(c) and Fig. \ref{fig:fig7}(c), pulsating flow of higher frequencies may require correspondingly higher concentration frequency to reduce the error. These results may indicate that the existence of a minimum concentration frequency for acquiring the smallest MAPE in SIV for a specific flow frequency.

\section{Discussion}
The conventional SIV method for extracting height-averaging fluid velocity from scalar signal transport in a shallow microfluidic channel using the integral minimization algorithm works well in a noise-free dataset of scalar field (Fig. \ref{fig:fig3}), however, it is very sensitive to noisy data of scalar field and has poor robustness ( Fig. \ref{fig:fig4}). It is for the purpose of overcoming the noise sensitivity and improving the robustness of conventional SIV method that DNN-SIV (Fig. \ref{fig:fig2}) is proposed and validated by simulation in this study. To the best of our knowledge, such DNN-SIV method has never been proposed in the past.\par

Conventional SIV method is effective on the premise that the measured scalar field is sufficiently resolved and relatively noise free, such that the spatial and temporal derivatives of the measured scalar field can be accurately evaluated. However, such noise requirements are difficult to be satisfied in the measurements of concentration field in microfluidic chips. In the turbulence velocimetry, improved SIV approach was proposed by Gillissen \citep{gillissen2018space}, that reconstructs not only the velocity field but also the scalar field. And the reconstruction error of SIV is remarkably reduced in turbulence generated by a falling soap film. Thus, we take a completely novel approach based on the deep learning to overcome the noise sensitivity of the conventional SIV method. The superiority of the DNN-SIV method in this problem is illustrated in Fig. \ref{fig:fig4} compared with the conventional SIV method. Besides, this method also demonstrates, such as advantages of stability, efficiency and robustness (Fig. \ref{fig:fig5} -- \ref{fig:fig7} and Table. \ref{tab:table2}).\par

The key point determining the accuracy of the velocity extraction from the scalar signal transport in a shallow straight microchannel is the scalar field itself without doubt. Obviously, if the concentration field in the microchannel does not vary with space and time, it is certain that none of velocity can be extracted from such scalar field with zero information entropy. That is why dynamic concentration condition is particularly adopted in the DNN-SIV. Hence, two fundamental issues about the scalar concentration field can be summarized. What is a ``good'' scalar field for velocity extraction and how to generate this scalar field in the microchannel.\par

For the first issue, we must refocus on the scalar transport equation which is composed of both convection and diffusion terms (Eq. \ref{Eq:4}). The diffusion term in scalar transport equation is dependent on the concentration gradient and independent of the flow velocity. Thus, information for velocity extraction is mainly enclosed in the convection term while retrieving velocity from the concentration field. This does not mean, of course, that the diffusion term has only negative effects. It contributes to the formation of the spatial concentration gradients in some cases. Thus, we conclude that the velocity can be extracted from the scalar field only if there is a spatiotemporal gradient of scalar concentration along the direction of velocity. It is not surprising to deduce that the more significant the spatiotemporal gradient of scalar field is, the more accurate the extracted velocity is (Fig. \ref{fig:fig5}). And absolute value of magnitude of the scalar concentration plays an ineffective role (Fig. \ref{fig:fig7}(a)).\par

For the second issue, the prerequisite condition is to design an appropriate dynamic concentration signal being input from inlet of microchannel. Yet, it needs to be emphasized that the scalar field in the microchannel is not determined by these dynamic concentration boundary conditions alone, but a combination of the input concentration signal and the flow characteristics. So, we investigated and analyzed the influence of the fluid flow characteristics and the scalar concentration signal transport in Sect \ref{Sec:ScalarFLuid}. It could be concluded from Fig. \ref{fig:fig6} and Fig. \ref{fig:fig7} that sophisticated design of scalar concentration signal can significantly reduce the MAPE of extracted velocity. It is instructive that the minimum error of extracted velocity may be achieved by setting an optimal concentration boundary condition at the entrance for specific flow characteristics in a microchannel chip.\par

The idea of DNN-SIV can extend to other shapes of microchannels with the high aspect ratio in which fluid flow can be simplified to a planner flow. An actual velocimetry system based on this idea is needed to be built for experiments to further investigate the technical specifications of the proposed method, such as measurement ranges, spatial and temporal resolution, error range and so forth. We have analyzed the effect of fluid flow characteristics and scalar concentration signal transport in Sect \ref{Sec:ScalarFLuid}, but the mathematical principles underlying it remain to be revealed. The collective influence of multiple factors needs to be further clarified to obtain the optimal concentration boundary conditions for a certain velocity profile analytically. In addition, despite the physics-informed neural network technique is adopted, the prediction performance of the trained network is still deficient while using different parameters of flow characteristics and input scalar concentration signal. But the issue of generalizability seems to be the most common problem in the application of neural networks. Inspired by the SIV theory, we believe that the proposed DNN-SIV can be further improved to have wide generalization as conventional SIV methods while maintaining its advantages.

\section{Conclusions}
SIV is an important flow visualization and measurement technology applied in dense velocity field and turbulent flow. It happens to be a very suitable technique for velocimetry of microfluidic chips with little modifications. We firstly introduce the concept of SIV into the measurement of laminar flow with small Reynolds Number in shallow microchannels, where diffusion is evident and convection is not dominant. Dynamic concentration boundary conditions are adopted at the inlet of the microchannel to improve the performance of velocity extraction. DNN-SIV coupled with scalar transport equation is also proposed in this paper to overcome the flaw of conventional SIV method and proven its advantages of stability and efficiency through numerical simulations. The velocimetry parameters of SIV have also been studied. We believe that this classical SIV concept will be revitalized in the measurement of fluid velocity in microfluidic chips with the application of deep learning technology.

\section*{Acknowledgment}
The research reported here was supported by grants from National Natural Science Foundation of China (No. 31971243 and No. 11802054), and the Fundamental Research Foundation for the Central Universities in China (No. DUT20YG113).

\bibliographystyle{unsrtnat}
\bibliography{Manuscript}

\begin{thebibliography}{54}
\providecommand{\natexlab}[1]{#1}
\providecommand{\url}[1]{\texttt{#1}}
\expandafter\ifx\csname urlstyle\endcsname\relax
  \providecommand{\doi}[1]{doi: #1}\else
  \providecommand{\doi}{doi: \begingroup \urlstyle{rm}\Url}\fi

\bibitem[Whitesides(2006)]{whitesides2006origins}
George~M Whitesides.
\newblock The origins and the future of microfluidics.
\newblock \emph{Nature}, 442\penalty0 (7101):\penalty0 368--373, 2006.

\bibitem[Stone et~al.(2004)Stone, Stroock, and Ajdari]{stone2004engineering}
Howard~A Stone, Abraham~D Stroock, and Armand Ajdari.
\newblock Engineering flows in small devices: microfluidics toward a
  lab-on-a-chip.
\newblock \emph{Annu. Rev. Fluid Mech.}, 36:\penalty0 381--411, 2004.

\bibitem[Nge et~al.(2013)Nge, Rogers, and Woolley]{nge2013advances}
Pamela~N Nge, Chad~I Rogers, and Adam~T Woolley.
\newblock Advances in microfluidic materials, functions, integration, and
  applications.
\newblock \emph{Chemical Reviews}, 113\penalty0 (4):\penalty0 2550--2583, 2013.

\bibitem[Halldorsson et~al.(2015)Halldorsson, Lucumi, G{\'o}mez-Sj{\"o}berg,
  and Fleming]{halldorsson2015advantages}
Skarphedinn Halldorsson, Edinson Lucumi, Rafael G{\'o}mez-Sj{\"o}berg, and
  Ronan~MT Fleming.
\newblock Advantages and challenges of microfluidic cell culture in
  polydimethylsiloxane devices.
\newblock \emph{Biosensors and Bioelectronics}, 63:\penalty0 218--231, 2015.

\bibitem[Sackmann et~al.(2014)Sackmann, Fulton, and Beebe]{sackmann2014present}
Eric~K Sackmann, Anna~L Fulton, and David~J Beebe.
\newblock The present and future role of microfluidics in biomedical research.
\newblock \emph{Nature}, 507\penalty0 (7491):\penalty0 181--189, 2014.

\bibitem[Shields~IV et~al.(2015)Shields~IV, Reyes, and
  L{\'o}pez]{shields2015microfluidic}
C~Wyatt Shields~IV, Catherine~D Reyes, and Gabriel~P L{\'o}pez.
\newblock Microfluidic cell sorting: a review of the advances in the separation
  of cells from debulking to rare cell isolation.
\newblock \emph{Lab on a Chip}, 15\penalty0 (5):\penalty0 1230--1249, 2015.

\bibitem[Esch et~al.(2015)Esch, Bahinski, and Huh]{esch2015organs}
Eric~W Esch, Anthony Bahinski, and Dongeun Huh.
\newblock Organs-on-chips at the frontiers of drug discovery.
\newblock \emph{Nature Reviews Drug Discovery}, 14\penalty0 (4):\penalty0
  248--260, 2015.

\bibitem[Wu et~al.(2020)Wu, Liu, Wang, Feng, Wu, Zhu, Wen, and
  Gong]{wu2020organ}
Qirui Wu, Jinfeng Liu, Xiaohong Wang, Lingyan Feng, Jinbo Wu, Xiaoli Zhu,
  Weijia Wen, and Xiuqing Gong.
\newblock Organ-on-a-chip: Recent breakthroughs and future prospects.
\newblock \emph{Biomedical Engineering Online}, 19\penalty0 (1):\penalty0
  1--19, 2020.

\bibitem[Wilmer et~al.(2016)Wilmer, Ng, Lanz, Vulto, Suter-Dick, and
  Masereeuw]{wilmer2016kidney}
Martijn~J Wilmer, Chee~Ping Ng, Henri{\"e}tte~L Lanz, Paul Vulto, Laura
  Suter-Dick, and Rosalinde Masereeuw.
\newblock Kidney-on-a-chip technology for drug-induced nephrotoxicity
  screening.
\newblock \emph{Trends in Biotechnology}, 34\penalty0 (2):\penalty0 156--170,
  2016.

\bibitem[Shang et~al.(2017)Shang, Cheng, and Zhao]{shang2017emerging}
Luoran Shang, Yao Cheng, and Yuanjin Zhao.
\newblock Emerging droplet microfluidics.
\newblock \emph{Chemical Reviews}, 117\penalty0 (12):\penalty0 7964--8040,
  2017.

\bibitem[Lam et~al.(2005)Lam, Chen, and Yang]{lam2005depthwise}
YC~Lam, X~Chen, and C~Yang.
\newblock Depthwise averaging approach to cross-stream mixing in a
  pressure-driven microchannel flow.
\newblock \emph{Microfluidics and Nanofluidics}, 1\penalty0 (3):\penalty0
  218--226, 2005.

\bibitem[Douf{\`e}ne et~al.(2019)Douf{\`e}ne, Tourn{\'e}-P{\'e}teilh, Etienne,
  and Aubert-Pou{\"e}ssel]{doufene2019microfluidic}
Koce{\"\i}la Douf{\`e}ne, Corine Tourn{\'e}-P{\'e}teilh, Pascal Etienne, and
  Anne Aubert-Pou{\"e}ssel.
\newblock Microfluidic systems for droplet generation in aqueous continuous
  phases: A focus review.
\newblock \emph{Langmuir}, 35\penalty0 (39):\penalty0 12597--12612, 2019.

\bibitem[Lane et~al.(2012)Lane, Jantzen, Carlon, Jamiolkowski, Grenet, Ley,
  Haseltine, Galinat, Lin, Allen, et~al.]{lane2012parallel}
Whitney~O Lane, Alexandra~E Jantzen, Tim~A Carlon, Ryan~M Jamiolkowski,
  Justin~E Grenet, Melissa~M Ley, Justin~M Haseltine, Lauren~J Galinat,
  Fu-Hsiung Lin, Jason~D Allen, et~al.
\newblock Parallel-plate flow chamber and continuous flow circuit to evaluate
  endothelial progenitor cells under laminar flow shear stress.
\newblock \emph{Journal of Visualized Experiments}, 17\penalty0 (59):\penalty0
  e3349, 2012.

\bibitem[Mehling and Tay(2014)]{mehling2014microfluidic}
Matthias Mehling and Sava{\c{s}} Tay.
\newblock Microfluidic cell culture.
\newblock \emph{Current Opinion in Biotechnology}, 25:\penalty0 95--102, 2014.

\bibitem[van~der Meer et~al.(2009)van~der Meer, Poot, Duits, Feijen, and
  Vermes]{van2009microfluidic}
A.~D. van~der Meer, A.~A. Poot, M.~H.~G. Duits, J.~Feijen, and I.~Vermes.
\newblock Microfluidic technology in vascular research.
\newblock \emph{Journal of Biomedicine and Biotechnology}, 2009:\penalty0
  1--10, 2009.

\bibitem[Nauman et~al.(1999)Nauman, Risic, Keaveny, and
  Satcher]{nauman1999quantitative}
Eric~A Nauman, Kurtis~J Risic, Tony~M Keaveny, and Robert~L Satcher.
\newblock Quantitative assessment of steady and pulsatile flow fields in a
  parallel plate flow chamber.
\newblock \emph{Annals of Biomedical Engineering}, 27\penalty0 (2):\penalty0
  194--199, 1999.

\bibitem[Bacabac et~al.(2005)Bacabac, Smit, Cowin, Van~Loon, Nieuwstadt,
  Heethaar, and Klein-Nulend]{bacabac2005dynamic}
Rommel~G Bacabac, Theo~H Smit, Stephen~C Cowin, Jack~JWA Van~Loon, Frans~TM
  Nieuwstadt, Rob Heethaar, and Jenneke Klein-Nulend.
\newblock Dynamic shear stress in parallel-plate flow chambers.
\newblock \emph{Journal of Biomechanics}, 38\penalty0 (1):\penalty0 159--167,
  2005.

\bibitem[Wong et~al.(2016)Wong, LLanos, Boroda, Rosenberg, and
  Rabbany]{wong2016parallel}
Andrew~K Wong, Pierre LLanos, Nickolas Boroda, Seth~R Rosenberg, and Sina~Y
  Rabbany.
\newblock A parallel-plate flow chamber for mechanical characterization of
  endothelial cells exposed to laminar shear stress.
\newblock \emph{Cellular and Molecular Bioengineering}, 9\penalty0
  (1):\penalty0 127--138, 2016.

\bibitem[Yang et~al.(2010)Yang, Xu, and Wang]{yang2010manipulation}
Chun-Guang Yang, Zhang-Run Xu, and Jian-Hua Wang.
\newblock Manipulation of droplets in microfluidic systems.
\newblock \emph{TrAC Trends in Analytical Chemistry}, 29\penalty0 (2):\penalty0
  141--157, 2010.

\bibitem[Ong et~al.(2008)Ong, Zhang, Du, and Fu]{ong2008fundamental}
Soon-Eng Ong, Sam Zhang, Hejun Du, and Yongqing Fu.
\newblock Fundamental principles and applications of microfluidic systems.
\newblock \emph{Front. Biosci}, 13\penalty0 (1):\penalty0 2757--2773, 2008.

\bibitem[Sinton(2004)]{sinton2004microscale}
David Sinton.
\newblock Microscale flow visualization.
\newblock \emph{Microfluidics and Nanofluidics}, 1\penalty0 (1):\penalty0
  2--21, 2004.

\bibitem[Williams et~al.(2010)Williams, Park, and
  Wereley]{williams2010advances}
Stuart~J Williams, Choongbae Park, and Steven~T Wereley.
\newblock Advances and applications on microfluidic velocimetry techniques.
\newblock \emph{Microfluidics and Nanofluidics}, 8\penalty0 (6):\penalty0
  709--726, 2010.

\bibitem[Wereley and Meinhart(2010)]{wereley2010recent}
Steven~T Wereley and Carl~D Meinhart.
\newblock Recent advances in micro-particle image velocimetry.
\newblock \emph{Annual Review of Fluid Mechanics}, 42:\penalty0 557--576, 2010.

\bibitem[Meinhart et~al.(1999)Meinhart, Wereley, and Santiago]{meinhart1999piv}
Carl~D Meinhart, Steve~T Wereley, and Juan~G Santiago.
\newblock {PIV} measurements of a microchannel flow.
\newblock \emph{Experiments in Fluids}, 27\penalty0 (5):\penalty0 414--419,
  1999.

\bibitem[Lindken et~al.(2009)Lindken, Rossi, Gro{\ss}e, and
  Westerweel]{lindken2009micro}
Ralph Lindken, Massimiliano Rossi, Sebastian Gro{\ss}e, and Jerry Westerweel.
\newblock Micro-particle image velocimetry ($\mu${PIV}): recent developments,
  applications, and guidelines.
\newblock \emph{Lab on a Chip}, 9\penalty0 (17):\penalty0 2551--2567, 2009.

\bibitem[Westerweel et~al.(2013)Westerweel, Elsinga, and
  Adrian]{westerweel2013particle}
Jerry Westerweel, Gerrit~E Elsinga, and Ronald~J Adrian.
\newblock Particle image velocimetry for complex and turbulent flows.
\newblock \emph{Annual Review of Fluid Mechanics}, 45\penalty0 (1):\penalty0
  409--436, 2013.

\bibitem[Adamczyk and Rimai(1988)]{adamczyk19882}
AA~Adamczyk and L~Rimai.
\newblock 2-dimensional particle tracking velocimetry ({PTV}): technique and
  image processing algorithms.
\newblock \emph{Experiments in Fluids}, 6\penalty0 (6):\penalty0 373--380,
  1988.

\bibitem[Kreizer et~al.(2010)Kreizer, Ratner, and Liberzon]{kreizer2010real}
Mark Kreizer, David Ratner, and Alex Liberzon.
\newblock Real-time image processing for particle tracking velocimetry.
\newblock \emph{Experiments in Fluids}, 48\penalty0 (1):\penalty0 105--110,
  2010.

\bibitem[Wang and Wang(2009)]{wang2009measurement}
Haoli Wang and Yuan Wang.
\newblock Measurement of water flow rate in microchannels based on the
  microfluidic particle image velocimetry.
\newblock \emph{Measurement}, 42\penalty0 (1):\penalty0 119--126, 2009.

\bibitem[Su and Dahm(1996{\natexlab{a}})]{su1996scalar1}
Lester~K Su and Werner~JA Dahm.
\newblock Scalar imaging velocimetry measurements of the velocity gradient
  tensor field in turbulent flows. i. assessment of errors.
\newblock \emph{Physics of Fluids}, 8\penalty0 (7):\penalty0 1869--1882,
  1996{\natexlab{a}}.

\bibitem[Su and Dahm(1996{\natexlab{b}})]{su1996scalar2}
Lester~K Su and Werner~JA Dahm.
\newblock Scalar imaging velocimetry measurements of the velocity gradient
  tensor field in turbulent flows. ii. experimental results.
\newblock \emph{Physics of Fluids}, 8\penalty0 (7):\penalty0 1883--1906,
  1996{\natexlab{b}}.

\bibitem[Chen et~al.(2015)Chen, Zill{\'e}, Shao, and Corpetti]{chen2015optical}
Xu~Chen, Pascal Zill{\'e}, Liang Shao, and Thomas Corpetti.
\newblock Optical flow for incompressible turbulence motion estimation.
\newblock \emph{Experiments in Fluids}, 56\penalty0 (1):\penalty0 1--14, 2015.

\bibitem[Kucukal et~al.(2021)Kucukal, Man, Gurkan, and
  Schmidt]{kucukal2021blood}
Erdem Kucukal, Yuncheng Man, Umut~A Gurkan, and BE~Schmidt.
\newblock Blood flow velocimetry in a microchannel during coagulation using
  particle image velocimetry and wavelet-based optical flow velocimetry.
\newblock \emph{Journal of Biomechanical Engineering}, 143\penalty0 (9), 2021.

\bibitem[Gillissen et~al.(2018)Gillissen, Vilquin, Kellay, Bouffanais, and
  Yue]{gillissen2018space}
Jurriaan~JJ Gillissen, Alexandre Vilquin, Hamid Kellay, Roland Bouffanais, and
  Dick~KP Yue.
\newblock A space--time integral minimisation method for the reconstruction of
  velocity fields from measured scalar fields.
\newblock \emph{Journal of Fluid Mechanics}, 854:\penalty0 348--366, 2018.

\bibitem[Heitz et~al.(2010)Heitz, M{\'e}min, and
  Schn{\"o}rr]{heitz2010variational}
Dominique Heitz, Etienne M{\'e}min, and Christoph Schn{\"o}rr.
\newblock Variational fluid flow measurements from image sequences: synopsis
  and perspectives.
\newblock \emph{Experiments in Fluids}, 48\penalty0 (3):\penalty0 369--393,
  2010.

\bibitem[Kalnay(2003)]{kalnay2003atmospheric}
Eugenia Kalnay.
\newblock \emph{Atmospheric modeling, data assimilation and predictability}.
\newblock Cambridge University Press, 2003.

\bibitem[Papadakis and M{\'e}min(2008)]{papadakis2008variational}
Nicolas Papadakis and {\'E}tienne M{\'e}min.
\newblock Variational assimilation of fluid motion from image sequence.
\newblock \emph{SIAM Journal on Imaging Sciences}, 1\penalty0 (4):\penalty0
  343--363, 2008.

\bibitem[Li et~al.(2013)Li, Li, Cao, and Qin]{li2013transport}
Yong-Jiang Li, Yizeng Li, Tun Cao, and Kai-Rong Qin.
\newblock Transport of dynamic biochemical signals in steady flow in a shallow
  y-shaped microfluidic channel: Effect of transverse diffusion and
  longitudinal dispersion.
\newblock \emph{Journal of Biomechanical Engineering}, 135\penalty0
  (12):\penalty0 121011, 2013.

\bibitem[Chen et~al.(2016)Chen, Gao, Zeng, Liu, Luan, and Qin]{chen2016shaped}
Zong-Zheng Chen, Zheng-Ming Gao, De-Pei Zeng, Bo~Liu, Yong Luan, and Kai-Rong
  Qin.
\newblock A y-shaped microfluidic device to study the combined effect of wall
  shear stress and atp signals on intracellular calcium dynamics in vascular
  endothelial cells.
\newblock \emph{Micromachines}, 7\penalty0 (11):\penalty0 213, 2016.

\bibitem[Li et~al.(2018)Li, Cao, and Qin]{li2018transmission}
Yong-Jiang Li, Tun Cao, and Kai-Rong Qin.
\newblock Transmission of dynamic biochemical signals in the shallow
  microfluidic channel: nonlinear modulation of the pulsatile flow.
\newblock \emph{Microfluidics and Nanofluidics}, 22\penalty0 (8):\penalty0
  1--13, 2018.

\bibitem[Burman et~al.(2020)Burman, Gillissen, and
  Oksanen]{burman2020stability}
Erik Burman, JJJ Gillissen, and Lauri Oksanen.
\newblock Stability estimate for scalar image velocimetry.
\newblock \emph{arXiv Preprint arXiv:2008.09451}, 2020.

\bibitem[Kou and Zhang(2021)]{kou2021data}
Jiaqing Kou and Weiwei Zhang.
\newblock Data-driven modeling for unsteady aerodynamics and aeroelasticity.
\newblock \emph{Progress in Aerospace Sciences}, 125:\penalty0 100725, 2021.

\bibitem[Bezgin et~al.(2021)Bezgin, Schmidt, and Adams]{bezgin2021data}
Deniz~A Bezgin, Steffen~J Schmidt, and Nikolaus~A Adams.
\newblock A data-driven physics-informed finite-volume scheme for nonclassical
  undercompressive shocks.
\newblock \emph{Journal of Computational Physics}, 437:\penalty0 110324, 2021.

\bibitem[Chen and Gu(2021)]{chen2021learning}
Chun-Teh Chen and Grace~X Gu.
\newblock Learning hidden elasticity with deep neural networks.
\newblock \emph{Proceedings of the National Academy of Sciences}, 118\penalty0
  (31), 2021.

\bibitem[Raissi et~al.(2020)Raissi, Yazdani, and Karniadakis]{raissi2020hidden}
Maziar Raissi, Alireza Yazdani, and George~Em Karniadakis.
\newblock Hidden fluid mechanics: Learning velocity and pressure fields from
  flow visualizations.
\newblock \emph{Science}, 367\penalty0 (6481):\penalty0 1026--1030, 2020.

\bibitem[Raissi et~al.(2019)Raissi, Perdikaris, and
  Karniadakis]{raissi2019physics}
Maziar Raissi, Paris Perdikaris, and George~E Karniadakis.
\newblock Physics-informed neural networks: A deep learning framework for
  solving forward and inverse problems involving nonlinear partial differential
  equations.
\newblock \emph{Journal of Computational Physics}, 378:\penalty0 686--707,
  2019.

\bibitem[Barwey et~al.(2019)Barwey, Hassanaly, Raman, and
  Steinberg]{barwey2019using}
Shivam Barwey, Malik Hassanaly, Venkat Raman, and Adam Steinberg.
\newblock Using machine learning to construct velocity fields from {OH-PLIF}
  images.
\newblock \emph{Combustion Science and Technology}, pages 1--24, 2019.

\bibitem[Cai et~al.(2021)Cai, Li, Zheng, Kong, Dao, Karniadakis, and
  Suresh]{cai2021artificial}
Shengze Cai, He~Li, Fuyin Zheng, Fang Kong, Ming Dao, George~Em Karniadakis,
  and Subra Suresh.
\newblock Artificial intelligence velocimetry and microaneurysm-on-a-chip for
  three-dimensional analysis of blood flow in physiology and disease.
\newblock \emph{Proceedings of the National Academy of Sciences}, 118\penalty0
  (13):\penalty0 e2100697118, 2021.

\bibitem[Feng et~al.(2007)Feng, Olsen, Hill, and Fox]{feng2007simultaneous}
Hua Feng, Michael~G Olsen, James~C Hill, and Rodney~O Fox.
\newblock Simultaneous velocity and concentration field measurements of
  passive-scalar mixing in a confined rectangular jet.
\newblock \emph{Experiments in Fluids}, 42\penalty0 (6):\penalty0 847--862,
  2007.

\bibitem[Corpetti et~al.(2009)Corpetti, H{\'e}as, M{\'e}min, and
  Papadakis]{corpetti2009pressure}
Thomas Corpetti, Patrick H{\'e}as, Etienne M{\'e}min, and Nicolas Papadakis.
\newblock Pressure image assimilation for atmospheric motion estimation.
\newblock \emph{Tellus A: Dynamic Meteorology and Oceanography}, 61\penalty0
  (1):\penalty0 160--178, 2009.

\bibitem[Wallace and Vukoslav{\v{c}}evi{\'c}(2010)]{wallace2010measurement}
James~M Wallace and Petar~V Vukoslav{\v{c}}evi{\'c}.
\newblock Measurement of the velocity gradient tensor in turbulent flows.
\newblock \emph{Annual Review of Fluid Mechanics}, 42:\penalty0 157--181, 2010.

\bibitem[He et~al.(2016)He, Zhang, Ren, and Sun]{he2016deep}
Kaiming He, Xiangyu Zhang, Shaoqing Ren, and Jian Sun.
\newblock Deep residual learning for image recognition.
\newblock In \emph{2016 IEEE Conference on Computer Vision and Pattern
  Recognition (CVPR)}, pages 770--778, 2016.

\bibitem[Glorot and Bengio(2010)]{glorot2010understanding}
Xavier Glorot and Yoshua Bengio.
\newblock Understanding the difficulty of training deep feedforward neural
  networks.
\newblock In \emph{Proceedings of the thirteenth international conference on
  artificial intelligence and statistics}, pages 249--256, 2010.

\bibitem[Kingma and Ba(2014)]{kingma2014adam}
Diederik~P Kingma and Jimmy Ba.
\newblock Adam: A method for stochastic optimization.
\newblock \emph{arXiv preprint arXiv:1412.6980}, 2014.

\end{thebibliography}

\end{document}